\documentclass{emulateapj}

\newcommand{\mdot}{$\dot M$}
\newcommand{\zpup}{$\zeta$~Pup}

\newcommand{\myemail}{jaz8@pitt.edu}

\slugcomment{To appear in ApJL}

\shorttitle{On the importance of interclump medium}
\shortauthors{Zsarg\'{o}et al.}

\begin{document}

\title{On the Importance of the Interclump Medium for Superionization:\\
        \ion{O}{6} Formation in the Wind of \zpup\ }

\author{J.~Zsarg\'{o}\altaffilmark{1,2}, D.~J.~Hillier\altaffilmark{2}, J.-C.~Bouret\altaffilmark{3},
T.~Lanz\altaffilmark{4}, M.~A.~Leutenegger\altaffilmark{5,6}, D.~Cohen\altaffilmark{7}}

\altaffiltext{1}{E-mail:~\myemail}

\altaffiltext{2}{Department of Physics and Astronomy, University of Pittsburgh, 3941 O'Hara St., 
Pittsburgh, PA 15260, USA}

\altaffiltext{3}{Laboratoire d'Astrophysique de Marseille, CNRS-Universit\'{e} de Provence, 
13388 Marseille cedex 13, FRANCE}

\altaffiltext{4}{Department of Astronomy, University of Maryland, College Park, MD~20742, USA}

\altaffiltext{5}{Goddard Space Flight Center, 8800 Greenbelt Road, Greenbelt, Maryland 20771, USA }

\altaffiltext{6}{NASA Postdoctoral Fellow}

\altaffiltext{7}{Department of Physics and Astronomy, Swarthmore College, 500 College Ave, Swarthmore, 
PA 19081, USA}

\begin{abstract}
We have studied superionization and X-ray line formation in the spectra of \zpup\
using our new stellar atmosphere code (XCMFGEN) that can be used to simultaneously analyze
optical, UV, and X-ray observations. 
Here, we present results on the formation of the \ion{O}{6}\ 
$\lambda\lambda 1032, 1038$ doublet.
Our simulations, supported by simple theoretical calculations, show
that clumped wind models that assume void in the interclump space 
cannot reproduce the observed \ion{O}{6} profiles. 
However, enough \ion{O}{6}\ can be produced if the voids are filled by a low density gas.
The recombination of \ion{O}{6} is very efficient in the dense material but in the tenuous interclump region an observable amount of \ion{O}{6} can be maintained.
We also find that different UV resonance lines are sensitive to different density
regimes in \zpup\ :
\ion{C}{4} is almost exclusively formed within the densest regions, while
the majority of \ion{O}{6} resides between clumps.
\ion{N}{5} is an intermediate case, with contributions from both the tenuous gas and clumps.
\end{abstract}

\keywords{stars:early-type -- stars:winds, outflow -- X-rays: stars -- X-rays:individual:\zpup}

\section{Introduction}

One of the surprising discoveries of the {\it Copernicus} satellite was the strong P-Cygnii 
profiles of superions, such as \ion{O}{6} and \ion{N}{5}, in the FUV spectra of many O and B stars 
\citep{sno76}.
The only viable explanation for the presence of \ion{O}{6} is Auger ionization by X-rays 
from \ion{O}{4} \citep{cas79} which is the dominant form of oxygen in many O-type stars.
The X-ray emission, necessary for Auger ionization, was later detected by the first X-ray telescopes 
\citep[e.g.,][]{har79, sew79}.

The origin of the stellar X-ray emission was another enigma until the ``wind-shock'' mechanism 
\citep{luc80}  became the accepted explanation.
Massive stars posses strong line-driven winds in which the material is accelerated by numerous
C, N, O, and Fe transitions \citep[see e.g.,][]{pau90, CAK}.
It was known from the conception of the line-driven wind theory that such flows are unstable and prone 
to the formation of dense clumps and shocks \citep[see e.g.,][]{owo91,luc80}.
The large-scale flow energy is converted to heat in the shock fronts producing high temperature plasma.
Numerical simulations confirm \citep{fel97, owo88} that at least the soft X-ray emission of early-type stars
can be explained by this mechanism.

Evidence for density inhomogeneities (or clumped winds) is provided by variability studies of both WR stars \cite[][and references therein]{lep99}, and O stars \cite[e.g.,][]{eve98, lep08}.
Further, density inhomogeneities allow the electron-scattering wings of emission lines to be reduced to the observed level while maintaining the strength of emission lines \citep{hil91, ham98, hil99}.
More recently, \cite{cro02} and \cite{hil03} found that they could not simultaneously fit the H$\alpha$  and 
\ion{P}{5} $\lambda\lambda$1120 profiles in normal O supergiants without assuming an inhomogeneous density distribution in the wind. 
Using a more statistical approach, \cite{mas03} showed that the phosphorus ionization structure was consistent with expectations only if lower than conventional mass-loss rates were used in their analysis
of \ion{P}{5} $\lambda\lambda$1120. 
Additional observational evidence for wind clumping come from {\it Chandra} and {\it XMM-Newton} high-resolution X-ray spectra of O stars. 
These spectra revealed that X-ray lines suffer less absorption in the wind than predicted by ``smooth" models \citep[e.g.,][and references therein]{kra03b,coh05,wal07}.

Superionization has received only limited attention since the work of \citeauthor{mac93} \citeyearpar{mac93, mac94}. 
The effect was introduced into modern stellar atmosphere codes \citep[e.g., {\it WM-basic},][]{pau01},
but we are unaware of any work that has revisited the question in light of 
the high-resolution X-ray observations, improved X-ray emission calculations and the new results on clumping.
With improvements to CMFGEN, we are developing tools and techniques to move towards this goal.
As part of this effort we discovered that the interclump medium is crucial to explain the 
\ion{O}{6} doublet profile in $\zeta$~Pup. 
In this paper we demonstrate the effect and discuss its implications.
In \S\ref{sec:obs} we briefly describe our code, the observations we used,
and our models. We present and discuss our results in \S\ref{sec:result}--\S\ref{sec:con}.       

\section{Observations, Tools, and Models} \label{sec:obs}

We coadded {\it Copernicus} U1 scans  for $\zeta$~Pup 
to create the observed profile in Figure~\ref{fig1}. 
All but two of these scans were blocked U1 scans that minimize stray-light as described in \cite{rog73a,rog73b}.
Our data reduction \citep[see,][]{zsa08} produced an \ion{O}{6} profile that strongly resemble 
those in \cite{mac93} and \cite{mor76}.

We have used XCMFGEN \citep{zsa08} to solve for the ionization balance and the non-LTE
level populations in our stellar models.
XCMFGEN is a new version of CMFGEN \citep{hil98} that can perform X-ray emission calculations
in addition to the original CMFGEN tasks. 
We use APEC \citep{smi01} and its accompanying database of X-ray cross sections (APED) to 
calculate non-LTE level populations and emissivities in the X-ray emitting plasma.

\cite{bou05, bou08} used CMFGEN to derive stellar parameters for seven O supergiants.
We used their values for \zpup~to construct the models listed in Table~\ref{tab1}. 
Our improved models specifically addressed  formation of \ion{O}{6} and X-ray lines \citep{zsa08}.
The X-ray emitting plasma was distributed in the wind and
its emissivity was characterized by three plasma temperatures. 
The parametrization included a volume filling factor ($f_X$) and an initial turn-on 
radius (R$_0$), as in \cite{owo01}.
The volume filling factor was constant beyond R$_0$ and was adjusted until
the prescribed L$_X$/L$_{BOL}$ was met (see Table~\ref{tab1}).
Note that L$_X$ is the X-ray luminosity attenuated by the wind and not
the intrinsic luminosity.

To treat clumping we follow the ``volume filling factor'' approach where the material is compressed
into a fraction of the available volume with void in between \citep[see e.g.,][]{hil99, ham98}.
The ratio of the volume filled with material (clumps) to the total is
\begin{equation}
f_{cl}(r) = f_{\infty} + (1 - f_{\infty}) \cdot \exp( -v(r)/v_{cl} ) \label{eq:cl}
\end{equation}
where $f_{\infty}$ and $v_{cl}$ are free parameters.
Eq.~\ref{eq:cl} is an ad-hoc formula, motivated by hydrodynamical simulations, and provides a 
smooth wind at low $v(r)$ velocities ($r \sim R_*$).

In our models, $f_X$ and $f_{cl}$ are two independent parameters despite
the fact that both are referred to as ``filling factors''.
The two should be related since both of them are the results of the wind instability. 
However, our understanding of the line-driven winds is too poor to formulate this relationship.

\begin{table}
\caption{Model Parameters (representative for \zpup). \label{tab1}}
\begin{tabular}{lcccr}
\tableline\tableline
                &    &    &    \\
{\bf Photosphere/Wind}  &    &    &    \\
 R$_*$\tablenotemark{a} & \multicolumn{4}{c}{19~R$_{\odot}$} \\
 T$_{eff}$           & \multicolumn{4}{c}{39,000~K} \\
 log~$g$ (cgs)         &   \multicolumn{4}{c}{3.6}    \\
 L$_*$               & \multicolumn{4}{c}{7.0$\times$10$^5$~L$_{\odot}$} \\
 v$_{turb}$          &      \multicolumn{4}{c}{12~km~s$^{-1}$}         \\
 v~sin~i             &      \multicolumn{4}{c}{240~km~s$^{-1}$}        \\
 v$_{\infty}$        &      \multicolumn{4}{c}{2300~km~s$^{-1}$}       \\
 $\beta$\tablenotemark{b} &      \multicolumn{4}{c}{0.9}         \\
   & \multicolumn{2}{c}{\it Smooth Wind} &  \multicolumn{2}{c}{\it Clumped Wind} \\
 $\dot M$    &  \multicolumn{2}{c}{7.6$\times$10$^{-6}$~M$_{\odot}$yr$^{-1}$} &
\multicolumn{2}{c}{1.7$\times$10$^{-6}$~M$_{\odot}$yr$^{-1}$} \\
 $f_{\infty}$ & \multicolumn{2}{c}{$\ldots$} & \multicolumn{2}{c}{0.05}         \\
 v$_{cl}$            & \multicolumn{2}{c}{$\ldots$} & \multicolumn{2}{c}{150~km~s$^{-1}$} \\
                     &    &    &    \\
{\bf Plasma}   &    &    &    \\
 R$_0$\tablenotemark{c}               &     \multicolumn{4}{c}{1.5~R$_*$}         \\
 L$_X$/L$_{BOL}$     &   \multicolumn{4}{c}{1.5$\times$10$^{-7}$} \\
 T$_{pl}$~1\tablenotemark{a}          &     \multicolumn{4}{c}{4.6$\times$10$^6$~K}         \\
 T$_{pl}$~2          &     \multicolumn{4}{c}{2.4$\times$10$^6$~K}         \\
 T$_{pl}$~3\tablenotemark{a}          &     \multicolumn{4}{c}{1.7$\times$10$^6$~K}         \\
\tableline
\end{tabular}
\tablenotetext{a}{Value is from \cite{hil93}.}
\tablenotetext{b}{Power of the CAK velocity law \citep{CAK}.}
\tablenotetext{c}{Value is from \cite{kra03b, leu06}.}
\tablecomments{Stellar parameters are from \cite{bou08}, unless noted otherwise.
Plasma parameters are from \cite{zsa08}, unless noted otherwise. For simplicity, the same
emission measure $EM_{X} = \int n_e n_H dV$ was assumed for each $T_{pl}$.}
\end{table}

\section{Results and Discussion} \label{sec:result}

\begin{figure}
\includegraphics[angle=270,scale=0.35]{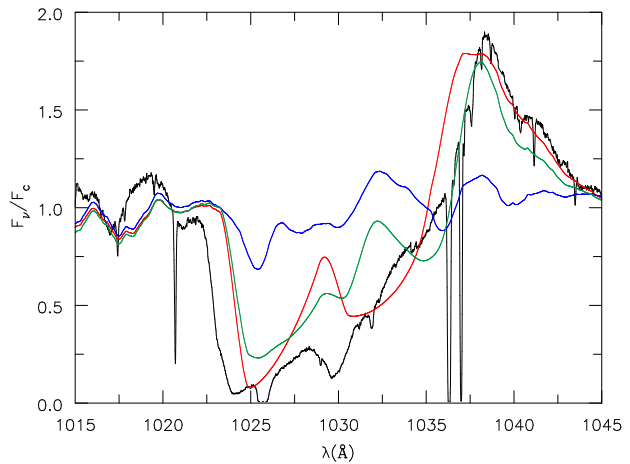}
\caption[]{Comparison of the {\it Copernicus} observation
of \ion{O}{6} $\lambda\lambda$1035 (rugged line) with those
calculated by XCMFGEN for a smooth wind (red), for a clumped wind 
model (blue), and for the interclump medium (green). 
See Table~\ref{tab1} and \S\ref{sec:result} for details.}
\label{fig1}
\end{figure}

Figure~\ref{fig1} shows our model calculations together with the {\it Copernicus}
spectrum.
Auger ionization produces strong \ion{O}{6} features in the case of the smooth wind model,
albeit a bit weaker than observed.
A better fit can be achieved by increasing the oxygen abundance and by assuming larger
turbulent velocities near $v_{\infty}$.
A detailed analysis of \zpup\ and the simultaneous fit to the X-ray and UV spectra, will be
presented in \cite{zsa08}.
The most important feature of Fig.~\ref{fig1} is the absence of an \ion{O}{6} profile
in the clumped wind model.
This cannot be rectified by fiddling with, for example, the oxygen abundance.
Adjustments at the order of magnitude level would be necessary to produce a visible \ion{O}{6} profile;
this indicates that something fundamental is wrong with the simplified ``clumped'' model.

There is a simple theoretical explanation for the weak \ion{O}{6} P-Cygnii profiles
in clumped models. 
Since the \ion{O}{6} $\lambda\lambda$1032, 1038 doublet forms only in the wind by 
redistribution of the stellar radiation, the optical depth controls the shape and 
strength of the profiles \citep[see e.g.,][]{lam87}.
In a spherical and accelerating flow it is appropriate to write the optical depth as
\begin{equation}
\tau_{O~{\rm VI}} \sim \frac{\pi e^2}{m c} \; \frac{f \, n_{O~{\rm VI}} \, \lambda_0 }{\frac{v}{r} + 
\left( \frac{dv}{dr} - \frac{v}{r} \right) \mu^2} 
\label{eq:tau}
\end{equation} 
where the atomic parameters have the usual meanings and $\arccos(\mu)$ is 
the angle between a line-of-sight and the radial direction.
The average \ion{O}{6} density can be written as
\begin{equation}
n_{O~{\rm VI}} = A_O \, n \, q_{O~{\rm VI}} \sim A_O \, \frac{\dot M}{4 \pi r^2 v(r) m_p } \, q_{O~{\rm VI}} \;
\label{eq:n_i} 
\end{equation} 
by using the oxygen abundance $A_O$, the \ion{O}{6} fraction $q_{O~{\rm VI}}$, and the average 
particle mass $m_p$.
Note, that $n_{O~{\rm VI}}$ is the {\it mean ion density over a Sobolev length}, so
\begin{equation}
n_{O~{\rm VI}} = n_{{O~{\rm VI}}, \, cl} \, f_{cl} = A_O \, n_{cl} \, f_{cl} \, q_{O~{\rm VI}}
\label{eq:n_icl} 
\end{equation} 
must be used if the wind material occupies only a fraction $f_{cl}$ of the available volume with 
$n_{cl}$ number density. 

The key unknown in the expression for $\tau_{O~{\rm VI}}$ is the \ion{O}{6} fraction. 
We can estimate $q_{O~{\rm VI}}$ by a simplified rate equation:
\begin{equation}
\frac{dn_{ O~{\rm VI} }}{dt} = n_{ O~{\rm IV} } \, \alpha (J_X) - n_{O~{\rm VI}} \, n_e \, \gamma = 0 ,
\label{eq:SEE}
\end{equation}
where $\alpha (J_X)$ is the X-ray-flux-dependent rate for Auger ionization, and $\gamma$ is 
the effective recombination rate coefficient.
Solving for the ion fraction,
\begin{equation}
\frac{n_{O~{\rm VI}}}{n_{O~{\rm IV}}} = \frac{q_{O~{\rm VI}}}{q_{O~{\rm IV}}} = \frac{\alpha 
(J_X)}{n_e \gamma}.
\label{eq:orat}
\end{equation}
Under the approximation that $q_{O~{\rm IV}} \sim 1$ and $n_e \sim n$, we find that $q_{O~{\rm VI}} =  
\frac{\alpha (J_X)}{n \gamma}$, or
\begin{equation}
n_{O~{\rm VI}} \; = \; A_O \, n \, q_{O~{\rm VI}} \; = \; A_O \, \frac{\alpha (J_X)}{\gamma}  \; .
\label{eq:nosix}
\end{equation}
The ion density of \ion{O}{6}, therefore, is {\it independent} of the wind density, and thus 
the mass-loss rate.
Substituting this in the expression for the optical depth, we find for a smooth wind
\begin{equation}
\tau_{O~{\rm VI}, sm} \sim  \frac{\pi e^2}{m c} \; \frac{A_O \, \alpha(J_X)}{\gamma} 
 \; \frac{f \, \lambda_0}{ \frac{v}{r} + \left( \frac{dv}{dr} - \frac{v}{r} \right) \mu^2} \; .
\label{eq:tauosix}
\end{equation}
The Sobolev optical depth in a smooth wind is thus also independent of the mass-loss rate.

Let us generalize this to a clumped wind. 
Equations~\ref{eq:orat} and \ref{eq:nosix} still give the \ion{O}{6} fraction and density, 
but these are the values for the dense clumps, and not means over a Sobolev length. 
Thus we need to use Eqs.~\ref{eq:tau}, \ref{eq:n_icl}, and \ref{eq:orat}, together with
the assumptions of $q_{O~{\rm IV}} \sim 1$ and $n_e \sim n$ to derive 
\begin{equation}
\tau_{O~{\rm VI},cl} \; \sim
\; f_{cl} \, \tau_{O~{\rm VI}, sm}
\label{eq:tauosixcl}
\end{equation}
in a clumped wind.

\begin{figure}
\includegraphics[angle=270,scale=0.35]{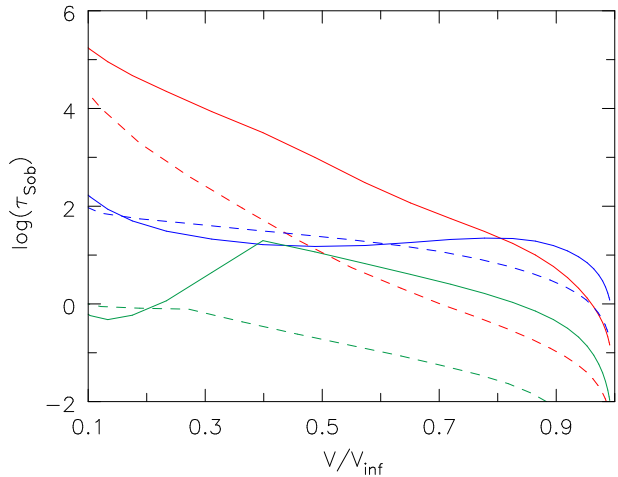}
\caption{
The radial optical depth as a function of normalized wind velocity for the 
stronger component of the \ion{N}{5} $\lambda\lambda$1240 (red), \ion{C}{4} 
$\lambda\lambda$1550 (blue), and \ion{O}{6} $\lambda\lambda$1035 (green) doublets.
The solid and dashed lines are for smooth and clumped wind models (Table~\ref{tab1}), 
respectively.
\label{fig2}}
\end{figure}

It is clear now why clumped models fail to produce observable \ion{O}{6} lines.
The volume filling factor for \zpup\ ($f_{cl}\sim$~0.05 at $r>>R_*$) 
decreases $\tau_{O~{\rm VI},cl}$ well below unity.
The results of our XCMFGEN simulations fully support the above calculation.
Figure~\ref{fig2} shows the radial optical depth of the important UV resonance lines in our
models (Table~\ref{tab1}).
The curves show what is predicted by Eq.~\ref{eq:tauosixcl} for small $f_{cl}$;
the \ion{O}{6} optical depth is nearly two orders of magnitude lower than in the smooth wind model.
The dense clumps, therefore, contribute little or nothing to the \ion{O}{6} profiles.

There are two other factors that further decrease the \ion{O}{6} optical depth, and hence 
adversely influence the model \ion{O}{6} profile.
First, \cite{bou08} found sub-solar oxygen abundances for all stars in their sample (a factor of
five difference for \zpup); this affects only comparisons with earlier calculations. 
Second, the X-ray flux available for Auger ionization tends to decrease when $f_{cl}$ is lowered;
the wind absorption is smaller, hence less X-ray emission is needed to meet the observed 
L$_X$/L$_{BOL}$.

What are the implications of our failure to produce observable \ion{O}{6}
lines with clumped winds?
Is the wind smoother than suggested by \cite{bou08}?
Their filling factor for \zpup~is not extremely low; it is very similar to those found for 
other O supergiants \citep{hil03,mas03}. Further, increasing $f_{cl}$  to 0.1 or 0.2 would not 
solve the problem.
It is more likely that the use of the  traditional ``filling factor'' approach in XCMFGEN causes
the failure to produce observable \ion{O}{6} lines.
Equations~\ref{eq:tauosix} and \ref{eq:tauosixcl} suggest that the part of the wind with the largest filling
factor will be the most important at producing O VI optical depth.
Thus we need to consider the influence of the tenuous interclump medium (ICM) on the \ion{O}{6} profile.

This is a non-trivial exercise; at least three components (dense clumps, ICM, and 
the hot plasma) need to be treated simultaneously for a fully self-consistent solution.
Fortunately, the radiation field at any radius is almost independent of the ICM.
Therefore, we can use $J_{\nu}(R)$ from our clumped model and solve the statistical 
equilibrium equations in the ICM only.
We created such a model from the clumped model of Table~\ref{tab1}.
We scaled down the densities by a factor of $f_{cl}^2$ to simulate the ICM at $r>>R_*$, and
imported the radiation field from the original clumped model.
We also assumed that the new model is smooth, reflecting the high volume filling fraction of 
the ICM.
Effectively, we took the ICM component out of the real (clump+ICM+hot plasma) wind and built a
stand-alone model for it.
Since we imported the radiation field from the clumped model, the coupling between the ICM and 
the rest of the wind has been taken into account at least in the first order. 

The choice of density in the ICM is somewhat arbitrary, but in the absence of reliable hydrodynamical 
predictions it is still reasonable.
There is a density contrast of four hundred between the dense clumps and the ICM at $r>>R_*$.
This means that the ICM contributes little to the overall mass-loss rates. 
If we combined the models for the ICM and clumps the total \mdot\ would be nearly identical 
(within 5\%) to that of the clumped model.

\begin{figure}
\includegraphics[angle=270,scale=0.35]{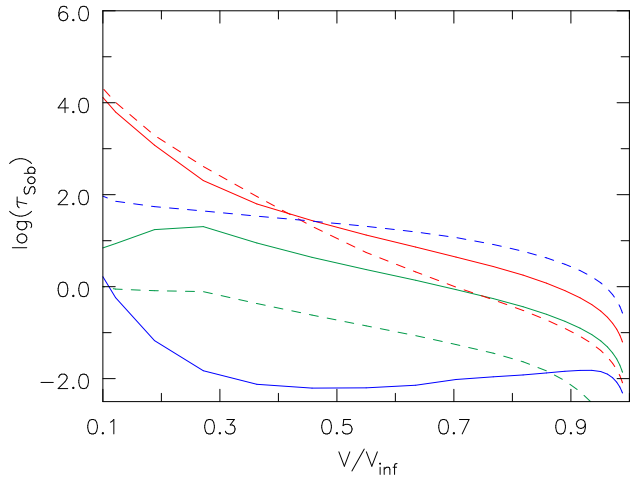}
\caption{
Same as figure~\ref{fig2} but the solid
curves are now for the interclump medium.
See \S\ref{sec:result} for details.  \label{fig3}}
\end{figure}

The result of this experiment is displayed in Figs.~\ref{fig1} and \ref{fig3}.
It is obvious from both figures that the ICM has substantial \ion{O}{6} optical depth 
and contributes a strong P-Cygnii profile. 
Our results also offer a glimpse on the behavior of other lines. 
The \ion{C}{4} optical depth in the ICM is very low and thus most of the \ion{C}{4} 
$\lambda\lambda$1550 lines are produced by the clumps.
The behavior of \ion{N}{5} $\lambda\lambda$1240 is very interesting.
It appears from Fig.~\ref{fig3} that both the ICM and the dense medium contribute. 

Our results warrant an investigation of the physical conditions in the ICM and also its role
in the formation of UV resonance lines in different environments (e.g., low and high density winds).  
For example, what is the density contrast between the ICM and clumps?
At very low densities
the average charge state of oxygen (and other species) may be drastically different 
from the one in the dense regions. 
Also, how closely does the ICM follow the velocity of the clumps?
Do \ion{O}{6} and \ion{C}{4} follow a different velocity structure? 
Further, how are the ICM and hot plasma related? 

\section{Conclusion} \label{sec:con}

In this letter we present our first results on superionization
in clumped winds, and showcase the potential of the interclump medium to
produce observable features. 
Clumped wind models that use the classical ``volume filling factor" approach 
(clumps with voids in between) cannot reproduce the observed \ion{O}{6} 
profile in $\zeta$~Pup. 
The recombination of \ion{O}{6} is too efficient and the necessary fractional 
abundance cannot be sustained in the clumps. 
However, a tenuous interclump medium can contribute enough \ion{O}{6} to produce an 
observable \ion{O}{6} profile. 
Only a small amount of mass is necessary in the interclump medium, so its overall 
effect on the derived mass-loss rates is negligible.

Our result highlights the need for improved treatment of clumping
in the winds of massive stars. 
It is impossible to achieve a simultaneous fit to all
UV P-Cygnii profiles with a single component wind model for \zpup.
Our simulations suggest that in $\zeta$~Pup, different UV resonance lines 
probe different density regimes. 
\ion{C}{4} is formed almost exclusively in the dense
material, while \ion{O}{6} likely originates from the interclump medium.
\ion{N}{5} is an intermediate case with similar contributions from both components.
In cooler O stars, when N$^{2+}$ become the dominant ionization stage, we might expect 
that \ion{N}{5}\ shows the same behavior as \ion{O}{6}\ in the hotter stars. 
Obviously other possible effects of the interclump medium, in both O and W-R stars, 
should be investigated.

\acknowledgments

This research was supported by STScI grant HST-AR-10693.02 and by SAO grant TM6-7003X.
We are also grateful to Dr. Randall Smith for providing us the source code of APEC, and
to the Chandra X-ray Center for the use of ATOMDB.
Maurice A. Leutenegger acknowledges support from a fellowship administered by Oak Ridge 
Associated Universities under the NASA Postdoctoral Program.
J.-C.~Bouret acknowledges financial support from the French National Research Agency (ANR) 
through program number ANR-06-BLAN-0105.

\end{document}